\documentclass[aps,prl,twocolumn,showpacs,floatfix,superscriptaddress,amssymb,
hamsmath]{revtex4}
\usepackage{graphicx}
\usepackage{dcolumn}
\usepackage{bm}
\usepackage{amsmath}

\usepackage{color}
\usepackage{dsfont}
\usepackage{wasysym}
\usepackage{ifsym}

\begin{document}
\title{Kondo screening regimes of a quantum dot with a single Mn ion}
\author{E. Vernek}
\affiliation{Instituto de F\'isica - Universidade Federal de Uberl\^andia -
Uberl\^andia, MG  38400-902 - Brazil}
\author{Fanyao Qu}
\affiliation{Instituto de F\'isica - Universidade de Bras\'ilia -
Bras\'ilia, DF  70919-970 - Brazil} 
\author{F. M. Souza}
\affiliation{Instituto de F\'isica - Universidade Federal de Uberl\^andia -
Uberl\^andia, MG  38400-902 - Brazil} 
\author{J. C. Egues}
\affiliation{Departamento de F\'isica e Inform\'atica, Instituto de F\'isica de S\~ao Carlos, Universidade de S\~ao Paulo,
13560-970 S\~ao Carlos, S\~ao Paulo, Brazil}
\author{E. V. Anda}
\affiliation{Departamento de F\'{\i}sica, Pontif\'{\i}cia
Universidade Cat\'olica, Rio de Janeiro-RJ, Brazil}

\date{\today}
\begin{abstract}
We study the Kondo and transport properties of a quantum dot with a single magnetic Mn 
ion connected to metallic leads. By employing a numerical renormalization group technique we show that depending on the value of ferromagnetic coupling strength between the local electronic spin and the magnetic
 moment of the Mn, two distinct Kondo regimes exist. In the weak coupling limit, the system
can be found in a completely screened Kondo state describing a local magnetic moment {\it decoupled} from the rest of the system. In contrast,
in the strong coupling regime the quantum dot  spin and the local magnetic moment form a single {\it large-spin} entity  partially Kondo screened.  A
crossover between these two regimes can be suitably tuned by
varying  the tunnel
coupling between the quantum dot electron and the leads. The model investigated here is also suitable to study magnetic molecules adsorbed on a metallic surface. The rich phenomenology of these systems is reflected in the conductance across the system.

\end{abstract}


\maketitle
Spin manipulation of localized impurities is of great interest in spintronics and quantum
computation\cite{spintronics}. In this context, diluted magnetic semiconductor quantum dots (DMSQDs) could play a prominent role as
they allow  the control of the spins of the  magnetic ions\cite{spincontrol,Divicenzo}. In general DMSQDs are grown in II-VI semiconductor composites with a few Mn atoms in each quantum dot (QD)\cite{Lee}. In these systems, the coupling between the
spins  of the electrons in the QD and those  of the manganese arises from the $sp-d$
exchange interaction.

More recently, the successful fabrication of QDs
doped 
with a single Mn$^{2+}$ ion\cite{Besombes,Kudelski,Maingault} has stimulated many optical and
transport    measurements\cite{Reiter,Gall,Goryca}, demanding a great deal of theoretical
efforts\cite{Govorov1,Govorov2,Astakhov,Besombes2}. 
Recent investigations of these systems have 
uncovered many interesting physical phenomena\cite{Besombes,Kudelski,Fanyao,Reiter,Gall,Goryca,Quem,Ribhu,Fanyao1}.
For instance, the exchange interaction makes single photon emitters active at
six different frequencies, thus serving as the basic framework
for the six-state qubit\cite{Govorov2}. In this context, a very exotic system composed of an 
``impurity'' with spin degrees of freedom coupled to a QD containing electrons
(the impurity is outside the QD) has been proposed and
studied, recently\cite{Ribhu}. Fewer theoretical works, however, have
addressed the transport properties in these systems\cite{Misiorny}.

\begin{figure}[h]
\centerline{\resizebox{2.0in}{!}{
\includegraphics{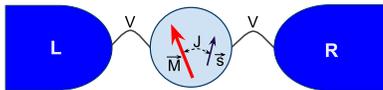}}}
\caption{\label{fig1}(color online) Single level quantum dot with a Mn ion . The 
large  arrow ($\vec{M}$) represents the spin of the Mn$^{2+}$ while the small arrow ($\vec s$) represents the spin of an electron in the. The matrix element $V$  allows for
electrons to hop -in and -off the quantum dot.}
\end{figure}

In this work we investigate the low temperature properties of  a quantum dot with a single magnetic 
Mn ion connected to leads. The study could be applied as well to analyze a magnetic molecules containing sites with correlated electrons, adsorbed on a metallic surface or connected to independent leads.  Although the ideas have this general scope,
to be concrete, we restrict our discussion to a system composed of a Mn$^{2+}$  ion implanted  in a small QD, coupled to two metallic (source and drain) leads,
schematically represented
in Fig.~\ref{fig1}.
It is well known that a QD connected to leads possesses a Kondo ground state similarly to what happens in magnetic impurities embedded  in
metals under temperature below the characteristic Kondo temperature ($T_K$)\cite{Manoharan}.
At the same time the electrons in the QD couple to the Mn$^{2+}$ magnetic moment
by a \emph{ferromagnetic}  exchange interaction, $J$, that can be optically or electrically tunned\cite{Besombes2,Fanyao}.
 The  antiferromagnetic case will be discussed 
in detail elsewhere\cite{antiferro}.   In our case, $T_K$
can be modified by tuning the hopping
matrix element $V$ that  connects the localized and the lead states, while
$J$  in turn can be tailored by properly choosing the size of the QD\cite{Fanyao1}. 
 
{Based on a numerical renormalization group (NRG) technique\cite{NRG}, our theoretical study shows two-distinct Kondo regimes: 1)
$T_K/|J|\gg 1$, where the QD spin is completely screened by the conduction spins comprising a Kondo state and the Mn$^{2+}$
is {\it decoupled} from the rest of the system and  2) $T_K/|J|\ll 1$, in which  the spins of 
the electrons in the
QD strongly couples to the Mn$^{2+}$ spin, forming a {\it large-spin} local
magnetic impurity that is partially screened by the conduction electrons: the
underscreened Kondo state}. A crossover between these two regimes
is achieved by suitably tuning the parameters of the system. 
Our system
 is described by the Hamiltonian $H=H_{imp}+ H_{bands}+H_{T}$
  where
\begin{eqnarray}
H_{imp}=\sum_\sigma
\varepsilon_{d\sigma}c^\dagger_{d\sigma}c_{d\sigma}+Un_{d\uparrow}n_{d\downarrow
}+J\mathbf{M}\cdot
\mathbf{s}
\end{eqnarray}
 corresponds to the single-level QD and the Mn$^{2+}$ ion, in which the operator $c^\dagger_{d\sigma}$ ($c_{d\sigma}$) creates (annihilates) an electron of spin
$\sigma$ 
with energy $\varepsilon_d$, $U$
is the local Coulomb repulsion and
$\mathbf{M}$ and $\mathbf{s}$ are the spin operators of the Mn$^{2+}$ ion and
of the QD, respectively. The Hamiltonian 
$H_{bands}=\sum_{\ell k\sigma}\varepsilon_{\ell
k}c^\dagger_{\ell k\sigma}c_{\ell k\sigma}
$ describes the conduction
band, where $c^\dagger_{\ell k\sigma}$ ($c_{\ell k\sigma}$) creates
(annihilates) an electron with momentum $k$, energy $\varepsilon_{\ell k}$ and
spin projection $\sigma$ in the $\ell$-th lead ($\ell=L,R$). The conduction bands are
characterized  by a constant density of
states  given by $\rho_0(\omega)=(1/2D)\Theta(D-|\omega|)$, where $D$ is the
half bandwidth and $\Theta(x)$ is the heavyside function. Finally,
$H_T=\sum_{\ell
k\sigma}V_{\ell}c^\dagger_{d\sigma}c_{\ell
k\sigma}+h.c.
$ describes the coupling between electrons in the QD and reservoirs.

Within  the NRG framework we are able to calculate
the relevant physical quantities, such as the entropy, the local magnetic moment and
the local retarded spectral function at the QD site. The later is necessary to calculate the 
zero-bias conductance.

Here  we mainly focus on investigating the
particle-hole symmetric case ($\varepsilon_d=-U/2$)  and normal leads, unless 
otherwise stated. We take $D$ as the energy unity. 
In order to illustrate clearly the underlying  physics of the numerical results presented latter,
we start off by briefly discussing a simplified model, i.~e., an  isolated impurity ($V=0$ in our model) which will help us to understand the results for the full system. In a single-electron
QD,
the spin
 is $s=1/2$. The coupling of the spin of the QD electron with the spin of the
Mn$^{2+}$ leads to a total angular momentum ${\mathbf L}={\mathbf
M}+{\mathbf s}$, with a total angular momentum quantum number, $l=|M-s|,\cdots,|M+s|$.
For the electron-Mn$^{2+}$ complex, the possible values of $l$ are $2$ and $3$. Assuming a ferromagnetic
coupling $J<0$, the ground  state has 
$l=3$ with degeneracy $2l+1=7$, corresponding to the
projections of the total angular momentum $\mathbf L$ along  the $z$-axis,  $l_z=-3,-2,...,2,3$. 

 We start our numerical analysis by studying
 the impurity contribution to the entropy and the magnetic moment, defined, respectively, as
 $ S(T)=S_t(T)-S_0(T)$
 and 

$\mu^2(T)=k_BT[\chi_t(T)-\chi_0(T)]$, where $\chi$ is the magnetic susceptibility.
 The subscripts $t$ and $0$ refer to the quantities calculated for the entire system and in the contribution of the conduction band alone, respectively.
We calculate these quantities within the usual NRG methodology\cite{NRGparameters}.
 Similarly, we calculate the spin-spin correlation ${\langle \textbf{M}\cdot \textbf{s} \rangle }$ an important quantity to characterize the system regime.

\begin{figure}[h]
\centerline{\resizebox{3.3in}{!}{
\includegraphics{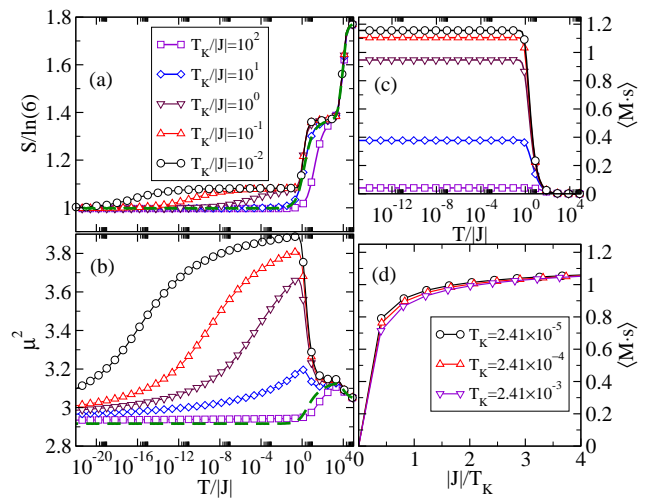}}}
\vskip0.25cm
\caption{\label{fig2-1}(color online) Entropy (a), magnetic moment (b) and spin-spin correlation as function (c)  as function of temperature for
various values of $T_K/|J|$ keeping $J=-2.0\times10^{-5}$. In the dashed (green) curve we use $J=0$ and $T_K=2.41\times10^{-4}$ and serves as reference. 
The panel (d) shows the spin-spin correlation as function of $J/T_K$ for various values of $T_K$. 
}
\end{figure}

In this paper  we  assume $U = 0.5$ and set $k_B=\mu_B=\hbar=1$.
  In Fig.~\ref{fig2-1}-a and
\ref{fig2-1}-b we show, respectively, the effects of the ferromagnetic exchange
interaction and temperature on the magnetic moment and entropy for $J=-2.0\times10^{-5}$ and
various values of $V$. Since the Kondo temperature $T_K$, evaluated in the absence of
ferromagnetic exchange interaction, is strongly dependent on V, and can be tuned
by changing $V$. As a reference,
the $J=0$ curve is also depicted (dashed green lines).

\paragraph{No exchange interaction.} Notice in Fig.~\ref{fig2-1}-a that at high temperature $T>\varepsilon_d$
$S\rightarrow 1.77Log(6)\approx Log(24)$, which indicates that there are
$6\times 4=24$ individual states,  $6$ from the Mn spin and $4$ from the dot, (spin and charge degrees of freedom), that can
 be thermally activated at that temperature. As the
temperature decreases, the entropy presents a plateau at $ Log (12)$, indicating that for $T\lesssim U/2$ the dot charge degrees of freedom are frozen. 
When the temperature decreases below $T_K$, the entropy for $J=0$  tends to $Log(6)$ (Fig.~\ref{fig2-1}-a), indicating that
the $6$ degenerate spin states of Mn is the only contribution to the entropy because the QD electron and conduction electrons are locked into a Kondo singlet. 

\paragraph{Finite ferromagnetic $J$ case.} Here the competition between
the exchange interaction energy $J$ and $T_K$  determines the
ground state of the system. The ground state is a Kondo singlet (KS) for 
$|J| \ll T_K$ with an uncoupled Mn ion and it becomes a local ferromagnetic state (LFS) as
$|J|\gg T_K$ due to the Mn-dot spin coupling that creates a {\it large spin} underscreened by the conduction electrons.  Here  we use the expression $T_K=\sqrt{\Gamma U}\exp{[-\pi U/8\Gamma]}$ to estimate the
Kondo temperature of the system in
the absence of the Mn atom (or $J=0$), with
$\Gamma=\pi V^2/D$ being the hybridization constant. In the intermediary case ($T_K\sim J$) the
system presents a crossover region between the regimes  mentioned above. In the following we analyze in
detail these several regimes.
 \paragraph*{Small-$J$ regime -- $|J|\ll T_K$.} In the regime of very weak exchange-interaction, when $T\ll T_K$, the regular full screened KS state emerges.  When $T < T_K$, as the
temperature decreases,  the KS ground state is formed by the strongly coupled dot and conduction electron spins.
 The characteristic of this regime is
clearly illustrated in the curve for  $T_K/|J|=10^{2}$ ($\square$) symbols
in Fig.~\ref{fig2-1}-a, where almost no plateau at $S=Log(7)$ is observed.The
entropy goes directly to the
 $S=Log(6)$ plateau. The screening of the QD-electron spin by the conduction electrons leaves the Mn-ion
free. Hence the magnetic moment of the system is only due to the 6-fold degenerate
state of the Mn atom, which at $T=0$ gives
$\mu^2=2[(-5/2)^2+(-3/2)^2+(-1/2)^2]/6=35/12\approx 2.92$, as clearly seen in
Fig.~\ref{fig2-1}-b ($\square$ curve).  In order to confirm this 
observation, in Fig.~\ref{fig2-1}-c we show  the spin-spin correlation
$\langle
\mathbf{M}\cdot \mathbf{s}\rangle$ as a function of the temperature using the same
parameters as in Fig.~\ref{fig2-1}-a. Notice that for a given $T_K/|J|$, as the temperature decreases, the
correlation rapidly increases, becoming  constant for $T\lesssim
|J|$, indicating the formation of a large effective localized spin. For $T_K/|J|=10^2$ the correlation
remains close to zero, thus indicating that the Mn ion is almost
fully
decoupled from the rest of the system.

\paragraph*{Large-$J$ regime -- $|J |\gg T_K$.}
In Figs.~\ref{fig2-1}-a and \ref{fig2-1}-b  this regime is better represented by
the curve for $T_K/|J|=10^{-2}$. In Fig.~\ref{fig2-1}-a we observe
that the entropy drops to a new plateau $\sim Log(7)$ for temperatures below
$|J|$ ($=2\times10^{-5}$ for this case). 
This new
value results from the ferromagnetic coupling between the $\mathbf{s}$
and $\mathbf{M}$. 
As we have discussed above, the enhancement of the magnetic moment corresponds to an unstable fix point, characterized by
7-fold degenerate state, corresponding to a total angular momentum $l=3$. In this
regime the QD and the Mn$^{2+}$ ion together comprise a large local
magnetic moment. This magnetic moment is, for sufficiently low temperature, partially  screened by the
 single channel conduction electron spins. This can be clearly seen in Fig.~\ref{fig2-1}-b, where the magnetic moment is shown as a function of T.  In this regime, $|J |\gg T_K$, the $\mu^2$ is suppressed by decreasing the temperature.
The local magnetic moment is only partially screened because there is only one conduction electron channel to screen the total spin. In the limit $T\rightarrow 0$ the system is undergoes into the underscreened Kondo regime, independently of the ratio $T_K/|J|$. This is reflected in the values $\mu^2\rightarrow 2.92$, $S \rightarrow Log(6)$ and $\langle\mathbf{M}\cdot \mathbf{s}\rangle\approx 1$, indicating a clear ferromagnetic correlation between the Mn and the dot spins.

\paragraph*{Intermediate regime -- $J \sim T_K$.} 
 In this case the system is in a crossover region between the two previous analyzed regimes. The amplitude of this region can be seen very clearly from  Fig.~\ref{fig2-1}-d where we show $\langle\mathbf{M}\cdot \mathbf{s}\rangle$  {\it vs}.  $|J|/T_K$ for three different values of
$T_K$. Notice that the correlation rapidly increases
 for $|J|/T_K\lesssim 1$ and saturates slowly for larger values of $|J|/T_K$, achieving
the value $5/4$ for $J/|T_K|\rightarrow \infty$. The region where the correlation changes rapidly corresponds in the parameter space to the crossover region. An inspection of the figure permits to conclude that the relevant parameter that controls the moment correlations is the quantity $J/T_K$ as we obtain the same universal function for the different values of $T_K$ taken.

The QD density of states, $\rho_d(\omega)=-\pi^{-1}{\tt
Im}[G^{r}_{dd}{(\omega)}]$, where $G^{r}_{dd}{(\omega)}$ is the Fourier  transform of the double time retarded Green's function, is calculated adopting standard NRG procedures. Within the same framework   the  zero-bias conductance $G$ across
the QD  is calculated using the Green function formalism with the Landauer-type formula\cite{Meir},
\begin{eqnarray}
G=\frac{2e^2}{h}\Gamma\int_{-\infty}^\infty{\tt
Im}[G^r_{
dd } (\omega)]\left[
-\partial f(\omega)/\partial \omega\right]d \omega,
\end{eqnarray}
where $f(\omega)$ is the Fermi function.
\begin{figure}[h]
\centerline{\resizebox{3.0in}{!}{
\includegraphics{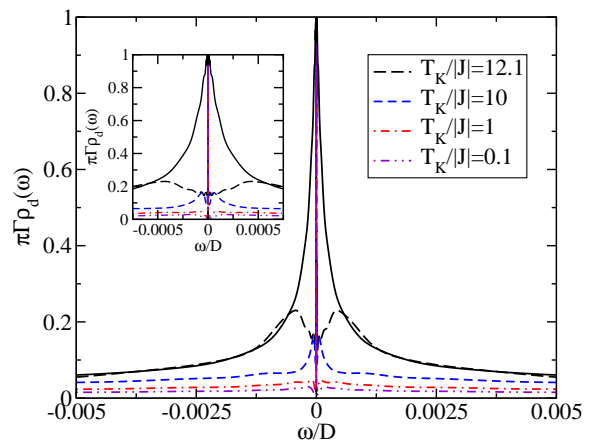}}}
\caption{\label{fig4-1}(color online) Spectral function {\it vs}. $\omega$ near the Fermi level for various values of $T_K/|J|$ showing how the Kondo peak is affected due to the coupling $J$. The line corresponds to $J=0$ and the same $T_K$ as in  the dashed (black) line. The inset shows a zoon-in around the Fermi level.}
\end{figure}

In Fig.~\ref{fig4-1} we show $\rho_{d}(\omega)$ for various values of $T_K/|J|$ near the Fermi level. Notice that even for $T_K>|J|$  (e.~g., $T_K/|J|=12.1$) line the Kondo peak split into a three-peak structure (including the very sharp peak at zero) due to the coupling $J$. The characteristic energy of this regime is given by the width of the complete three peak structure.
 When compared to the solid curve for $J=0$ we see that the change in the density of states is restricted to the Kondo peak ($\sim T_K$) region and there is a clear collapse of the two curves for $|\omega| \gtrsim T_K$. 
  As $T_K/|J|$ decreases we see a dramatic distortion in the Kondo peak: the three-peak structure tends to disappear and essentially the density of states is dominated by the central sharp peak at the Fermi level. It is interesting to observe that despite the strong modification of the spectral function, the height of the peak  \emph{at}  the Fermi level   remains $1/\pi\Gamma$, as predicted by the Friedel sum rule. As a result, no effect would be expected  for the conductance at $T=0$. However, as we discuss below, the presence of this central peak has essential consequences on the interesting behavior of the conductance as the temperature is increased. 

\begin{figure}[h]
\vskip0.25cm
\centerline{\resizebox{3.0in}{!}{
\includegraphics{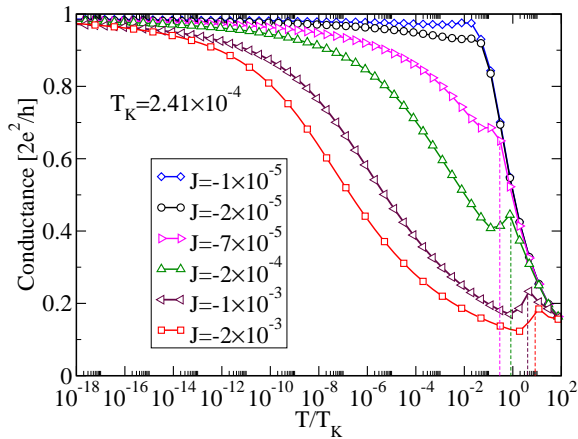}}}
\caption{\label{fig4-2}(color online) Conductance as function of temperature for various values of $J$ and fixed $T_K=2.41\times10^{-4}$. The vertical dashed lines indicate the position of the various $|J|$ along the $x$-axis showing a good coincidence with the spikes observed in the conductance.}
\end{figure}

 In Fig.~\ref{fig4-2} we show the conductance as function of temperature for various values of $J$. We notice two distinct regimes: i) for $J\ll T_K$ the conductance drops at $T\sim T_K$, where the effect of the temperature is to take the system out of the standard Kondo regime and ii) for $|J|\gg T_K$ the conductance drops for much lower temperature. For intermediate values of $J$, such as in $|J|=7\times10^{-7}<T_K$, we observe that the behavior corresponding to the two regimes is contained in the same curve. The shape of the conductance differs completely whether $T<|J|$ or $T>|J|$. The appearance of these two regimes with temperature can be understood as following: by decreasing $T$, in the interval $|J|<T<T_K$, the system restores the behavior of the completely screened Kondo state at that temperature, as thermal excitations destroys the dot-magnetic atom spin-spin correlation (see Fig~\ref{fig2-1}-c). However, reducing $T$, after a crossover region, when $T<J$, {the dot-magnetic atom spin degrees of freedom are coupled and the system enters an underscreened Kondo state, characterized by a conductance that is significantly dependent upon temperature.}

In summary, we have investigated the Kondo regime of a system composed of a single  Mn$^{2+}$ ion 
in a QD coupled to metallic leads. Our numerical approach shows two distinct low temperature
regimes,  depending on how the ferromagnetic 
exchange interaction $J$  between the electrons in the QD and the magnetic moment of the Mn$^{2+}$ 
compares with $T_K$. In the weak regime ($T_K\gg|J|$), the QD  is locked in a Kondo state singlet
while the magnetic moment of the Mn$^{2+}$ decouples from the rest of the system, while in the
strong coupling regime ($T_K\ll J$), the QD and the Mn$^{2+}$ forms a spin $l=3$ impurity that couples
to the conduction band. In this case,  the impurity is underscreened by the Kondo correlation with
the conduction band electrons. 
From the experimental point of view, Fig.~\ref{fig4-2} shows that the nature of the Kondo regime is reflected very significantly on the conductance permitting, through a transport measurement, to fully characterize the spin configuration of the system with very interesting quantum information implications.
We expect that our results will stimulate experimental studies of these systems and contribute to the understanding of Kondo effect and transport in magnetic quantum dots.

We would like to thank the Brazilian agencies CNPq, CAPES, FAPERJ, FAPEMIG and FAPESP for financial support.


\begin{thebibliography}{apssamp}
\bibitem{antiferro} $J>0$  is also of interest and  is left for a future investigation. See also Ref.~\onlinecite{Misiorny}.
\bibitem{NRGparameters} We use a NRG discretization parameter $\Lambda=2.5$ and keep  2,000 of states at each iteration, before accounting for the degeneracies.
\end{thebibliography}
\end{document}